\documentclass{article}
\usepackage{xcolor} 
\usepackage{amsfonts}
\usepackage{amsmath}
\usepackage{amssymb}
\usepackage{graphicx}
\usepackage{systeme} 
\usepackage[utf8]{inputenc}
\usepackage{multirow} 
\usepackage{booktabs} 
\usepackage[english]{babel}
\usepackage{amsmath}
\usepackage{amsfonts}
\usepackage{amssymb}
\usepackage{graphicx}
\usepackage[small,bf]{caption} 
\usepackage[left=3cm,right=3cm,top=2cm,bottom=3cm]{geometry}
\usepackage{framed} 
\usepackage{color} 
\usepackage{wrapfig}\definecolor{shadecolor}{RGB}{220,220,220} 

\author{Jorge Pinochet and Giorgio Sonnino}
\title{\textbf{Entropy, Gravity, and an Apparent Violation of the Second Law}}
\begin{document}

\author{Jorge Pinochet$^{1}$, Giorgio Sonnino$^{2}$\\ \\
\small{$^{1}$\textit{Centro de Investigación en Educación (CIE-UMCE),}}\\
  \small{\textit{Universidad Metropolitana de Ciencias de la Educación,}}\\
 \small{\textit{Av. José Pedro Alessandri 774, Ñuñoa, Santiago, Chile.}}\\
\small{$^{2}$\textit{Department of Physics, Université Libre de Bruxelles (U.L.B.),}}\\
\small{\textit{Campus de la Plaine C.P. 224, Bvd du Triomphe, 1050 Brussels, Belgium.}}\\
\small{e-mail: jorge.pinochet@umce.cl\ ; e-mail: giorgio.sonnino@ulb.be}}

\date{}
\maketitle

\begin{abstract}
\noindent An interesting question to explore in physics classes is: Does gravity violate the second law of thermodynamics? If we turn to a general physics textbook for an answer, we will find little to no reference to the relationship between entropy and gravity. The same is often true for specialized textbooks. The goal of this work is to address this question by analyzing the behavior of an ideal gas in two simple scenarios: one where gravity is negligible and another where its effects play a significant role. We have shown that while gravity-driven systems may exhibit counterintuitive behaviors - such as local ordering through structure formation - the second law of thermodynamics remains valid when considering the entire system, including all emitted energy and radiation. Given the educational focus of this work and the complexity of the entropy–gravity relationship, we will omit detailed calculations that are not strictly necessary and instead focus on the simplest physical scenarios, analyzing four representative examples through simple calculations: the Sun, the limit of extreme contraction in black holes, the protostellar contraction sequence, and core collapse with neutrino cooling.


\end{abstract}

\section{Introduction}

A commonly used example in general physics textbooks to illustrate entropy is the free expansion of an ideal gas. This is an irreversible process in which a gas, composed of non-interacting point particles, expands inside an empty, isolated container. As the gas spreads out, its distribution becomes increasingly uniform, leading to an increase in entropy, which is consistent with the second law of thermodynamics. This law states that the entropy of an isolated system never decreases. \\

The standard description of free expansion assumes that, due to the small mass of the gas, gravitational effects can be ignored. However, if we were to increase the mass of the gas significantly so that the gravitational attraction between its particles becomes relevant, we would observe a different outcome: instead of expanding uniformly, the gas would contract, forming clumps and reducing its overall uniformity. This raises an intriguing question: Does this contraction imply a decrease in entropy? More broadly, does gravity violate the second law of thermodynamics? \\

This work addresses such questions by examining the behavior of ideal gases both with and without gravity, to assess the validity of the second law in each case. Although these topics are rarely discussed in standard physics textbooks [1,2,3,4,5,6,7], they offer a valuable opportunity to strengthen physical intuition and bridge concepts from thermodynamics and classical mechanics. This integrative approach can enrich second or third-year courses in Thermodynamics or Statistical Physics. With a pedagogical focus, this paper avoids lengthy derivations and instead emphasizes heuristic reasoning supported by simple, schematic figures. The arguments presented require only basic knowledge of calculus and introductory mechanics, making the material accessible to most undergraduate students in physics, engineering, or other science programs. Instructors may also find this content suitable for elective seminars, particularly the section exploring black-hole thermodynamics, which introduces students to a frontier topic in an approachable manner. \\

The article has been written without ever losing sight of the fact that it is aimed at the widest possible audience, composed mainly of physics educators and their students. Throughout the text, we have placed special emphasis on clarity and simplicity, avoiding any technicality that is not strictly necessary and reducing calculations to a minimum, but without sacrificing the rigor or accuracy required of a scientific work.

\section{Joule (Free) Expansion from Thermodynamics and Statistical Mechanics}\label{FG}

In this section, we shall treat the textbook ideal-gas Joule (free) expansion from both the thermodynamic and statistical points of view, show the step-by-step math, and explain why the entropy increases even though no heat or work is exchanged.

\subsection{Setup and assumptions}

\begin{figure}[h]
\center{
\includegraphics[width=9cm]{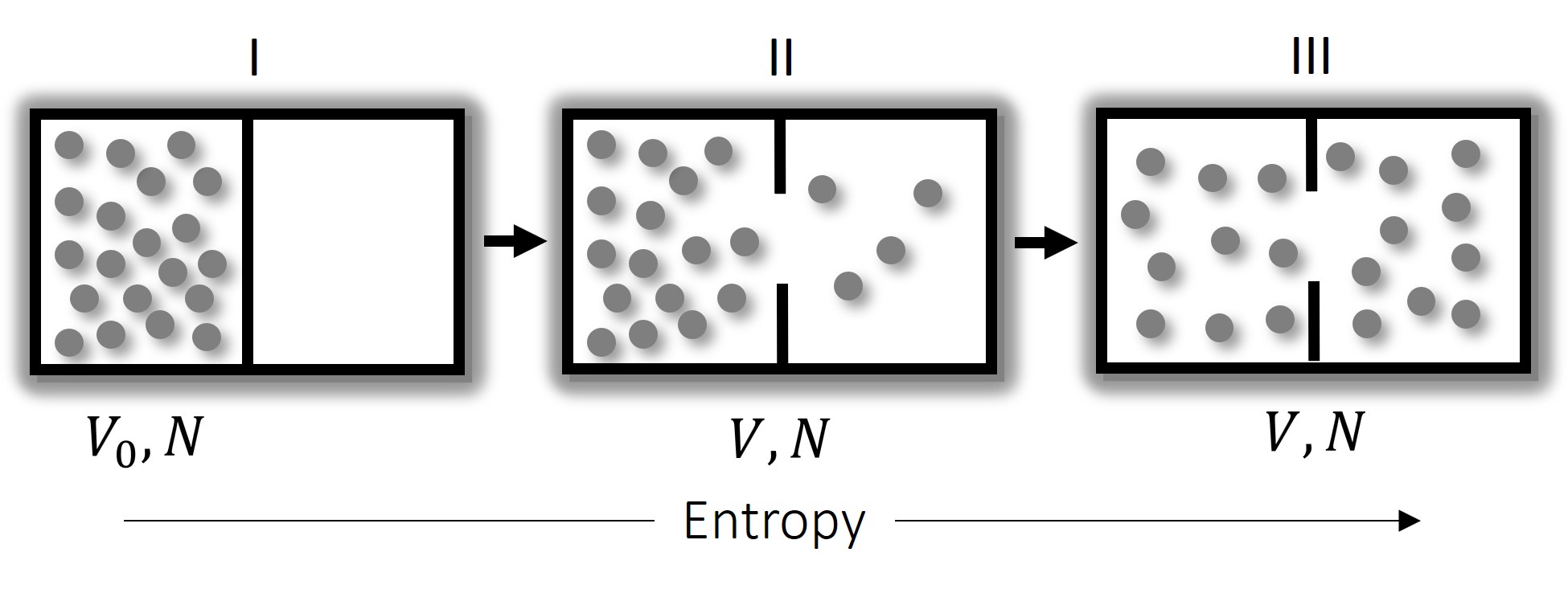}
\caption {Free expansion of an ideal gas. Entropy increases as the gas expands.}
\label{Fig1}}
\end{figure}

\noindent Let us consider an ideal monatomic gas (or any ideal gas) initially occupying a volume $V_i$ at temperature $T$. The gas is separated from an empty chamber by a removable partition. The whole vessel is thermally isolated (no heat exchange with the environment). At $t = 0$ the partition is removed, and the gas freely expands to fill volume $V_f$, with $V_f > V_i$ (see Figure~\ref{Fig1}).

\noindent Since the gas is ideal, the internal energy $U$ depends only on temperature: $U = U(T)$, and there are no interparticle potentials. No external work is done (expansion occurs in a vacuum), no heat flows, we have:
\begin{equation}\label{FG1}
\delta Q = 0, \qquad \delta W = 0
\end{equation}
\noindent Energy balance for the isolated gas reads:
\begin{equation}\label{FG2}
\Delta U = Q - W = 0 - 0 = 0
\end{equation}
\noindent  For an ideal gas $U = U(T)$, so $\Delta U = 0 \Rightarrow \Delta T = 0$. The temperature remains $T$ before and after expansion. The entropy change of the gas, $\Delta S$, between the initial and final equilibrium states can be computed by choosing any reversible path connecting them (entropy is a state function). Choose a reversible isothermal expansion at temperature $T$ from $V_i$ to $V_f$. For a reversible isothermal process:
\begin{equation}\label{FG3}
\delta Q_{\text{rev}} = P\, dV
\end{equation}
\noindent Using the ideal-gas law $P = nRT/V$
\begin{equation}\label{FG4}
\Delta S = \int_i^f \frac{\delta Q_{\text{rev}}}{T} = \int_{V_i}^{V_f} \frac{P\, dV}{T} = \int_{V_i}^{V_f} \frac{nR}{V}\, dV = nR \ln\frac{V_f}{V_i}
\end{equation}
\noindent Equivalently, in particle notation \(N\) and Boltzmann constant \(k_B\) (where \(R = N_A k_B\) and \(n = N/N_A\)):
\begin{equation}\label{FG5}
\Delta S = N k_B \ln\frac{V_f}{V_i} = nR \ln\frac{V_f}{V_i}
\end{equation}
\noindent Let us consider a numerical example for one mole ($n = 1$). $R = 8.31446261815324\, \mathrm{J\,mol^{-1}\,K^{-1}}$ and if \(V_f / V_i = 2\) (then $\ln 2 \simeq 0.693147$), we obtain:
\begin{equation}\label{FG7}
\Delta S = 8.3144626 \times 0.693147 = 5.763\, \mathrm{J/K}
\end{equation}
\noindent Thus, for one mole doubling its volume by free expansion, $\Delta S \approx 5.763\, \mathrm{J/K}$.Even though the process is adiabatic (\(Q = 0\)) and no work is done, the entropy of the isolated gas increases because the final macrostate has more accessible microstates (larger volume).

\subsection{Statistical (microscopic) derivations}

\noindent It is instructive to re-derive the above results by statistical microscopy. We shall prove this by following three different short routes.

\subsubsection{A. Counting-box/Boltzmann argument}

\noindent If we coarse-grain the container into identical spatial cells, the number of microstates where \(N\) particles occupy volume \(V\) scales like \(V^N\) (classically). Then Boltzmann entropy:
\begin{equation}\label{FG8}
S = k_B \ln \Omega \quad \Rightarrow \quad
\Delta S = k_B \ln\frac{\Omega_f}{\Omega_i}= k_B \ln\frac{V_f^N}{V_i^N} = N k_B \ln\frac{V_f}{V_i}
\end{equation}
\noindent i.e., we get the same result.

\subsubsection{B. Canonical partition function}
\noindent For a monatomic ideal gas, the single-particle thermal wavelength is:
\begin{equation}\label{FG9}
\lambda = \frac{h}{\sqrt{2\pi m k_B T}}
\end{equation}
\noindent The canonical partition function for \(N\) indistinguishable particles reads:
\begin{equation}\label{FG710}
Z(T,V,N) = \frac{1}{N!} \left( \frac{V}{\lambda^3} \right)^N
\end{equation}
\noindent The Helmholtz free energy is
\begin{equation}\label{FG711}
F = -k_B T \ln Z = -k_B T \big[ N\ln V - 3N \ln \lambda - \ln N! \big]
\end{equation}
\noindent and the entropy expression can be derived by differentiation:
\begin{equation}\label{FG12}
S = -\left(\frac{\partial F}{\partial T}\right)_V
\end{equation}
\noindent where only $\ln \lambda$ depends on $T$. Since $\lambda = h / \sqrt{2\pi m k_B T}$,
\begin{equation}\label{FG13}
\frac{\partial \ln \lambda}{\partial T} = -\frac{1}{2T}
\end{equation}
\noindent Differentiating,
\begin{equation}\label{FG14}
\frac{\partial F}{\partial T}
= -k_B \big[N\ln V - 3N\ln\lambda - \ln N!\big]- k_B T \big[-3N\, \frac{\partial \ln\lambda}{\partial T}\big]
\end{equation}
\noindent we obtain:
\begin{equation}\label{FG15}
S = k_B \big[N\ln V - 3N\ln\lambda - \tfrac{3}{2}N + \ln N!\big]
\end{equation}
\noindent Using Stirling’s approximation for large $N$:
\begin{equation}\label{FG16}
\ln N! \approx N\ln N - N + \tfrac{1}{2}\ln(2\pi N)
\end{equation}
\noindent we find:
\begin{equation}\label{FG17}
\frac{S}{k_B} = N\ln\frac{V}{N} - 3N\ln\lambda + \tfrac{5}{2}N + \tfrac{1}{2}\ln(2\pi N)
\end{equation}
\noindent Neglecting the small term \(\tfrac{1}{2}\ln(2\pi N)\) we finally get:
\begin{equation}\label{FG18}
\boxed{S = N k_B \left[ \ln\!\left(\frac{V}{N\lambda^3}\right) + \frac{5}{2} \right]}
\end{equation}
\noindent This is the Sackur-Tetrode entropy for a monatomic ideal gas \cite{schroeder} (see Appendix A for a more rigorous derivation of this expression).

\subsubsection{C. Microcanonical multiplicity}
\noindent For an ideal gas in the microcanonical ensemble, the multiplicity $\Omega$ (phase-space volume consistent with energy $E$) has a factor \(V^N\) (from position integration). If \(E\) is constant during free expansion, then
\begin{equation}\label{FG19}
\frac{\Omega_f}{\Omega_i} = \left(\frac{V_f}{V_i}\right)^N
\end{equation}
\noindent so again
\begin{equation}\label{FG20}
\Delta S = k_B \ln\frac{\Omega_f}{\Omega_i} = N k_B \ln\frac{V_f}{V_i}
\end{equation}
\noindent Let us clarify why entropy increases even though Liouville’s theorem preserves phase-space density. The fine-grained Gibbs entropy is expressed as
\begin{equation}\label{FG21}
S_{\mathrm{Gibbs}} = -k_B \int \rho(\Gamma, t) \ln \rho(\Gamma, t)\, d\Gamma
\end{equation}
\noindent which is constant under exact Hamiltonian evolution (Liouville’s theorem). This constancy is not in conflict with the observed increase in entropy. Boltzmann or coarse-grained entropy counts the phase-space volume compatible with a macrostate (e.g., \textit{all particles confined to the left half} vs.\ \textit{particles occupy the whole box}). Removing the partition increases the accessible macrostate volume. The fine-grained density filaments spread into the larger region and, under any realistic coarse-graining (finite measurement resolution), the effective distribution becomes uniform over the new volume. The coarse-grained entropy increases. Irreversibility arises from the special low-entropy initial condition and coarse-graining.
\vskip0.15truecm 
\noindent {\bf Physical interpretation of the obtained results and final remarks}

\noindent Free expansion increases entropy because the number of accessible microstates increases (spatial freedom grows). The process is irreversible: there is no spontaneous return to the original confined state without work or information input. For an isolated ideal gas:
\begin{equation}\label{FG22}
\boxed{\Delta S = nR \ln\!\left(\frac{V_f}{V_i}\right)}
\end{equation}
\noindent The increase is independent of the gas details (as long as it is ideal and $U=U(T)$); it depends only on particle number and volume ratio. This is the physics for the case “turn off gravity”. It establishes a baseline: expansion in a vacuum increases entropy. Contraction, corresponding to the gravitational case, requires additional accounting (work/energy/radiation) to avoid violating the second law. This will be the subject of the next section.

\section{Radiative Accounting in Gravitational Contraction}\label{WithG}

In this section, we give a self-contained, detailed, and quantitative account of the role of energy and radiation during the \emph{gravitational contraction} of an ideal gas. The central question is: when a gas contracts under its own gravity (so that its internal entropy tends to decrease), where is the compensating entropy that preserves the second law? The short answer is: the compensating entropy is carried away by the emitted energy (photons, neutrinos, gravitational waves, etc.).\\

\begin{figure}[h]
\center{
\includegraphics[width=4cm]{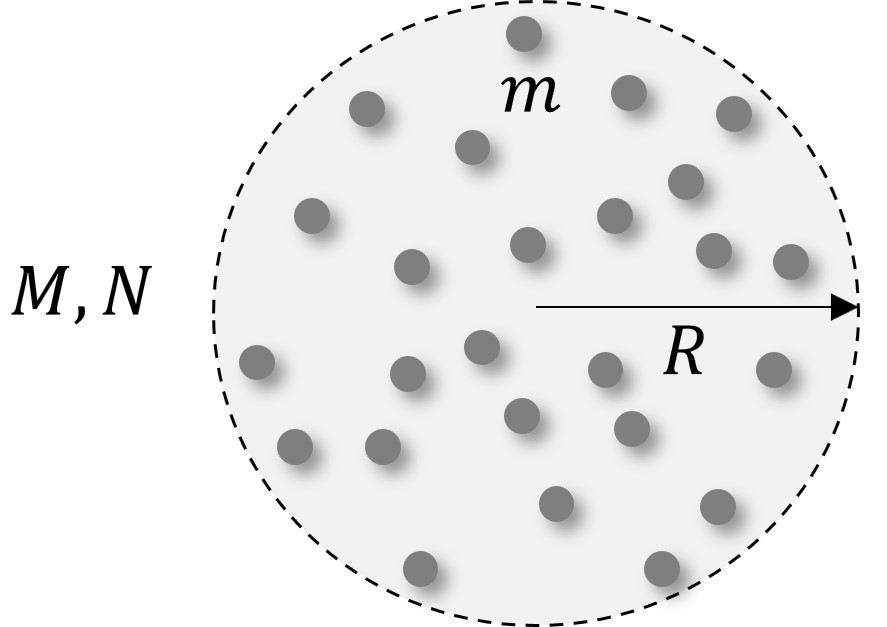}
\caption {A mass M of ideal gas with a spherical shape, composed of N particles of mass m that move randomly and are held together by the attraction they exert on each other.}
\label{Fig2}}
\end{figure}

We will demonstrate this explicitly by writing down the energy conservation law, expressing the rate of entropy change of the gas and the radiation, and combining them to obtain a manifestly non-negative total entropy production under the standard physical assumptions (quasi-static contraction and radiation leaving the system). We then discuss limiting regimes, numerical estimates (stellar mass), and the black hole limit (Bekenstein-Hawking entropy and the generalized second law). Consider a self-gravitating system composed of \(N\) identical particles of mass \(m\), total mass$M = Nm$, and characteristic radius $R(t)$ (see Figure 2). \\

\begin{figure}[h]
\center{
\includegraphics[width=9cm]{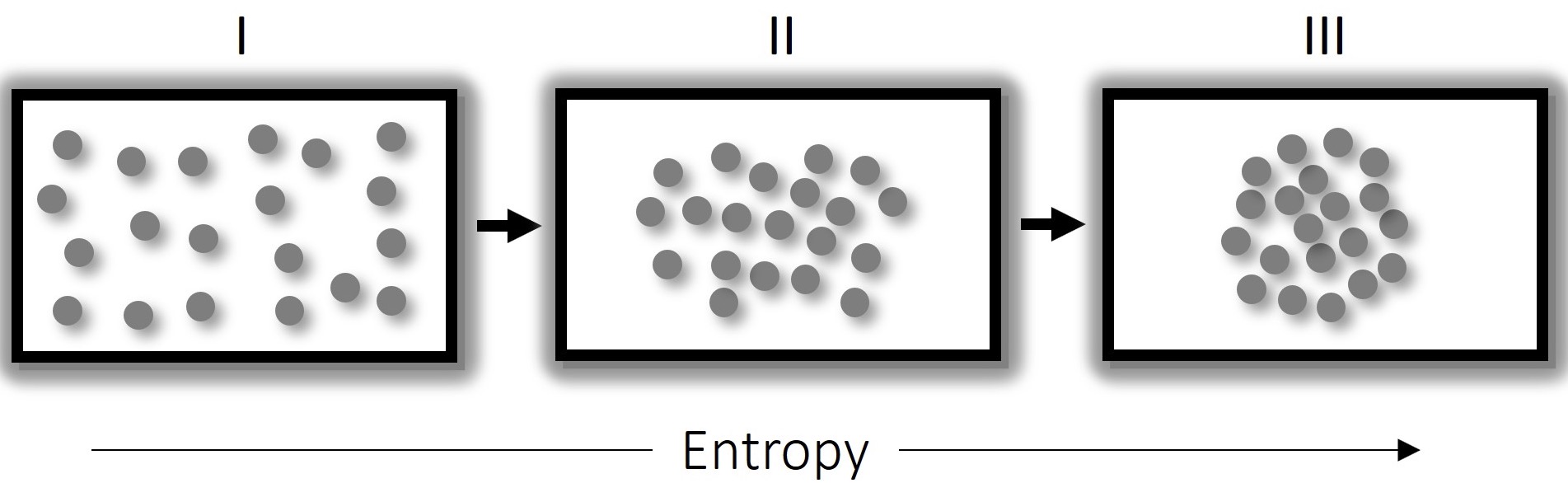}
\caption {Free contraction of an ideal gas. The increase in entropy is associated with a reduction in the volume and uniformity of the gas distribution.}
\label{Fig3}}
\end{figure}

\noindent For simplicity, take the system approximately spherical with instantaneous volume
\begin{equation}\label{WG1}
V(t)=\frac{4\pi}{3}R(t)^3
\end{equation}
\noindent The total (mechanical) energy is
\begin{equation}\label{eq:E_tot}
E(t)=K(t)+U(t)
\end{equation}
\noindent where $K$ is the total kinetic energy and $U$ the gravitational potential energy. For a roughly uniform sphere, one may write \cite{kittel}
\begin{equation}\label{eq:U_def}
U(t) = -\alpha\,\frac{G M^2}{R(t)},\qquad 
\alpha = 
\begin{cases} 
3/5 & \text{(uniform sphere)}\\[4pt] 
\mathcal O(1) & \text{(general bound configurations)} 
\end{cases}
\end{equation}
\noindent If the system is approximately virialized (or evolves through quasi-static sequences of near-equilibria), the scalar virial theorem for a gravitationally bound system gives \cite{maoz} (for an exhaustive justification, refer to Appendices~B and B1):
\begin{equation}\label{eq:virial}
2K + U = 0 \quad \Rightarrow \quad K = -\frac{U}{2}= \frac{\alpha G M^2}{2R}
\end{equation}
\noindent Identifying the kinetic energy with the thermal energy of the gas,
\begin{equation}\label{eq:K_T}
K = \frac{3}{2} N k_B T
\end{equation}
\noindent we obtain a characteristic temperature as a function of radius:
\begin{equation}\label{eq:T_of_R}
T(R) \;=\; \frac{\alpha G M m}{3 k_B R}
\end{equation}
\noindent Note the inverse dependence on $R$: a contraction ($R\downarrow)$ implies heating ($T\uparrow$). This is the origin of \emph{negative heat capacity} in self-gravitating systems. Figure~\ref{Fig3} illustrates the situation.
\noindent If the system radiates energy at a total power (luminosity) \(L(t)\) that leaves the system permanently, energy conservation reads
\begin{equation}\label{eq:energy_cons}
\frac{dE}{dt} = -L(t) + (\text{work / mass flux terms})
\end{equation}
\noindent where the parentheses indicate possible additional terms if mass is lost or external work is done. For the isolated situation with no mass loss and no external work, one can take
\begin{equation}\label{WG2}
\frac{dE}{dt} = -L(t)
\end{equation}
\noindent Using \(E=K+U\) and the virial relation, one may relate \(dE/dt\) to the contraction rate \(\dot R\). For instance, with \(U = -\alpha GM^2/R\) we have:
\begin{equation}\label{WG3}
\frac{dU}{dt} = \alpha G M^2 \frac{\dot R}{R^2}, \qquad \frac{dK}{dt} = -\tfrac12 \frac{dU}{dt} = -\tfrac12\alpha G M^2 \frac{\dot R}{R^2}
\end{equation}
\noindent Hence,
\begin{equation}\label{WG3}
\frac{dE}{dt} = \frac{d(K+U)}{dt} = -\tfrac12 \alpha G M^2 \frac{\dot R}{R^2} + \alpha G M^2 \frac{\dot R}{R^2} = \tfrac12 \alpha G M^2 \frac{\dot R}{R^2}
\end{equation}
\noindent Consistency with \(\dfrac{dE}{dt}=-L\) implies \(\dot R<0\) corresponds to \(L>0\) (energy leaves the system as it contracts). Now, we split the total entropy change into two contributions: the entropy change of the gas \(S_{\rm gas}\) and the entropy carried away by radiation \(S_{\rm rad}\). The total rate is
\begin{equation}\label{eq:Sdot_total}
\dot S_{\rm tot} \;\equiv\; \dot S_{\rm gas} + \dot S_{\rm rad}
\end{equation}
\noindent We compute now the expression for the gas entropy rate by assuming the so-called \textit{quasi-static approximation}. In other words, we assume the contraction proceeds sufficiently slowly that the gas can be treated as locally in thermodynamic equilibrium at temperature \(T_{\rm gas}(t)\). Then, the infinitesimal reversible heat absorbed by the gas reads:
\begin{equation}\label{WG4}
\delta Q_{\rm gas} = dE + P\, dV
\end{equation}
\noindent Here, $E$ is the internal/total mechanical energy of the gas used in the first law; signs follow the convention $\delta Q = dE + P\,dV$ for the system. Hence, for quasi-static evolution,
\begin{equation}\label{eq:Sdot_gas_def}
\dot S_{\rm gas} = \frac{1}{T_{\rm gas}} \left(\frac{dE}{dt} + P\frac{dV}{dt}\right)
\end{equation}
\noindent Note that during contraction \(dV/dt<0\) and (for a normal gas) \(P>0\), so the \(P dV\) term tends to make \( \dot S_{\rm gas}<0\) (the gas becomes more ordered). If the system emits radiation at (total) luminosity \(L(t)\), that radiation carries entropy away. For a \emph{general} spectrum, the instantaneous entropy flux (rate) associated with emitted energy can be written formally as
\begin{equation}\label{WG5}
\dot S_{\rm rad} \;=\; \int \frac{\mathrm{d}\dot E_\nu}{\langle \epsilon_\nu\rangle}
\end{equation}
\noindent where \(\mathrm{d}\dot E_\nu\) is spectral power and \(\langle\epsilon_\nu\rangle\) is an appropriate mean photon energy; this is not always convenient. A robust and simple approximation is to treat the escaping radiation as approximately thermal (blackbody) at some effective temperature \(T_{\rm rad}\) characteristic of the photosphere/emission region. For blackbody radiation, the relationship between energy and entropy densities yields a simple formula for the entropy emission rate:
\begin{equation}\label{eq:Sdot_bb}
\dot S_{\rm rad} \approx \frac{4}{3}\frac{L}{T_{\rm rad}} \quad\text{(blackbody emission)}
\end{equation}
\vskip0.15truecm 
\noindent To help students appreciate the validity of Eq.~(\ref{eq:Sdot_bb}), let us reason as follows. For equilibrium radiation, we have
\begin{equation}\label{WG6}
u = a T^4,\qquad s = \frac{4}{3}a T^3
\end{equation}
with $u$ and $s$ denoting the energy per unit volume and the entropy per unit volume, respectively. So, $s/u = 4/(3T)$. If energy \(L\) leaves the system per unit time, the associated entropy flux is \(\dot S \simeq (s/u)\,L = (4/3)L/T\). For non-thermal spectra, the numerical prefactor can differ; it is conservative to write
\begin{equation}\label{eq:Sdot_rad_general}
\dot S_{\rm rad} = \eta\,\frac{L}{T_{\rm rad}},\qquad 1 \lesssim \eta \lesssim \tfrac{4}{3}
\end{equation}
\noindent where \(\eta\) encodes spectral details (for late stages or neutrino-dominated cooling \(\eta\) will differ).  Combining \eqref{eq:Sdot_gas_def} and \eqref{eq:Sdot_rad_general} we get:
\begin{equation}\label{eq:Sdot_combined}
\dot S_{\rm tot} \;=\; \frac{1}{T_{\rm gas}}\Big(\frac{dE}{dt} + P\frac{dV}{dt}\Big) + \eta\,\frac{L}{T_{\rm rad}}
\end{equation}
\noindent Using \(dE/dt=-L\) (energy carried away by radiation), this becomes:
\begin{equation}\label{eq:Sdot_key}
\boxed{ \quad \dot S_{\rm tot} \;=\; L\Big(\frac{\eta}{T_{\rm rad}} - \frac{1}{T_{\rm gas}}\Big) + \frac{P\,\dot V}{T_{\rm gas}} \quad }
\end{equation}
\noindent where $\dot V=dV/dt<0$ during contraction.
\vskip0.15truecm
\noindent{\bf Physical Interpretation of Eq.~(\ref{eq:Sdot_key})}.

\noindent Let us interpret the single terms from the physical point of view:
\begin{itemize}
\item [$-$] The first term, \(L(\eta/T_{\rm rad}-1/T_{\rm gas})\), is the entropy balance between emitted radiation and the energy extracted from the gas. If the escaping radiation is cooler (lower mean energy per photon) than the gas, i.e.\ \(T_{\rm rad}<T_{\rm gas}\), the factor \(\eta/T_{\rm rad}\) can exceed \(1/T_{\rm gas}\) and the term is positive. This is the usual case: energy flows from hot gas to cooler radiation, producing entropy.
\item [$-$] The second term, \(P\dot V/T_{\rm gas}\), is negative during contraction and represents the decrease of gas entropy due to volume reduction (ordering). For the total entropy to be non-negative, the positive radiation term must compensate for this negative term.
\end{itemize}

\begin{figure}[h]
\center{
\includegraphics[width=8cm]{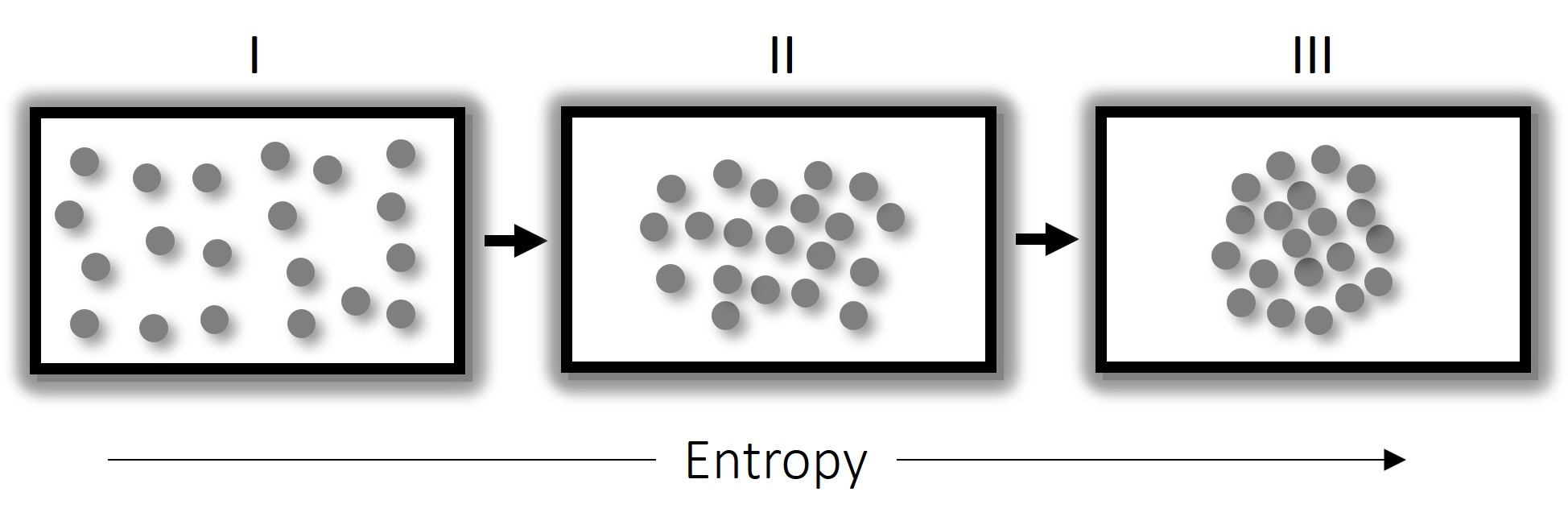}
\caption {By combining the rate of entropy change of the gas and the radiation, we get a non-negative total entropy production under the standard physical assumptions (quasi-static contraction and radiation leaving the system). The compensating entropy is carried away by the emitted energy (photons, neutrinos, gravitational waves, etc.).}
\label{Fig4}}
\end{figure}

\noindent Thus, the radiative luminosity $L$ and its spectrum (via $T_{\rm rad}$ and $\eta$) are the missing pieces that ensure the second law is satisfied. Figure~\ref{Fig4} illustrates the situation.

\section{Simple estimates and physical timescales}

It is helpful to estimate orders of magnitude to see that in realistic gravitational contraction, the radiative entropy easily overcompensates the gas entropy loss.

\subsection{Kelvin - Helmholtz (KH) timescale and typical luminosity}

\noindent We introduce the gravitational binding energy scale (order of magnitude)
\begin{equation}\label{WG7}
E_{\rm grav} \sim \frac{G M^2}{R}
\end{equation}
\noindent If the system radiates this energy on a timescale \(t_{\rm KH}\) (Kelvin--Helmholtz timescale), the typical luminosity is
\begin{equation}\label{eq:L_KH}
L \sim \frac{E_{\rm grav}}{t_{\rm KH}} \sim \frac{G M^2}{R\,t_{\rm KH}}
\end{equation}
\noindent For the Sun (\(M_\odot\), \(R_\odot\)), taking the observed luminosity \(L_\odot\) yields \(t_{\rm KH}\sim 3\times 10^7\) yr, which is consistent with the simple estimate \(E_{\rm grav}/L_\odot\).

\subsection{Entropy rates (order of magnitude)}
\noindent Let us use \eqref{eq:Sdot_key} and the blackbody approximation \(\eta = 4/3\):
\begin{equation}\label{WG8}
\dot S_{\rm tot} \approx L\Big(\frac{4}{3T_{\rm rad}} - \frac{1}{T_{\rm gas}}\Big) + \frac{P\dot V}{T_{\rm gas}}
\end{equation}
\noindent If the emission region has temperature \(T_{\rm rad}\) significantly lower than the interior gas temperature \(T_{\rm gas}\) (typical in optically thick objects where the photosphere is cooler than deep interior), the first term dominates and \(\dot S_{\rm tot}>0\).
Even when \(T_{\rm rad}\) is comparable to \(T_{\rm gas}\), the blackbody factor \(4/3\) helps make the radiation entropy production larger than the energy loss alone would suggest. 

\vskip0.15truecm
\noindent {\bf Concrete examples of entropy bookkeeping during a contracting episode}

\noindent It is instructive to illustrate the procedure by examining concrete examples of entropy accounting in gravitational contraction. Below, we report four relevant cases.

\noindent {\bf A) The entropy change in the sun}

\noindent Let us consider the sun-scale numbers. We take \(M=M_\odot\approx 2.0\times10^{30}\,\mathrm{kg}\). A rough estimate provides
\begin{equation}\label{WG9}
E_{\rm grav}\sim \frac{G M^2}{R_\odot}\approx \frac{6.67\times10^{-11}\cdot (2\times10^{30})^2}{7\times10^8} \sim 4\times10^{41}\ \mathrm{J}
\end{equation}
\noindent With present solar luminosity \(L_\odot\sim 3.8\times10^{26}\ \mathrm{W}\), the KH time is
\begin{equation}\label{WG10}
t_{\rm KH}\sim \frac{E_{\rm grav}}{L_\odot}\sim \frac{4\times10^{41}}{3.8\times10^{26}} \approx 1\times10^{15}\ \mathrm{s} \sim 3\times10^7\ \text{yr}
\end{equation}
While gravitational contraction (K-H mechanism) yields a characteristic timescale of order $10^7$ years, the actual age of the Sun ($\sim~ 4.6\times 10^9$ years) is much larger. This discrepancy is resolved by nuclear fusion: in the core of the Sun, hydrogen is converted into helium via the proton-proton chain, releasing energy of order $\sim 0.7\%mc^2$. This provides a long-term, steady energy source, allowing hydrostatic and thermal equilibrium over billions of years. Gravitational contraction remains relevant only in early (protostellar) phases and late evolutionary stages. The entropy flux in radiation at effective temperature \(T_{\rm rad}\approx 5800\ \mathrm{K}\) (solar photosphere) is approximately
\begin{equation}\label{WG11}
\dot S_{\rm rad} \approx \frac{4}{3}\frac{L_\odot}{T_{\rm rad}} \sim \frac{4}{3}\frac{3.8\times10^{26}}{5.8\times10^3} \sim 8.7\times10^{22}\ \mathrm{J/K\,s^{-1}}\approx 6.3\times10^{45}\ k_B\,\mathrm{s^{-1}}
\end{equation}
\noindent If one divides by Boltzmann's constant to give a dimensionless number of \(k_B\) per second, equivalently, keep units J/K/s. The gas interior is much hotter in the deep interior; the key is that radiation emitted from cooler layers has large entropy per unit energy compared to the interior energy per particle. Consider now an episode where the gas radius decreases from \(R_i\) to \(R_f\), with total radiated energy
\begin{equation}\label{WG12}
E_{\rm rad} = \int_{t_i}^{t_f} L(t)\,dt = E(R_i)-E(R_f) = \Delta E < 0
\end{equation}
\noindent The total entropy change is
\begin{equation}\label{WG13}
\Delta S_{\rm tot} = \Delta S_{\rm gas} + \Delta S_{\rm rad} = \int_{t_i}^{t_f} \frac{dE + P\,dV}{T_{\rm gas}} + \int_{t_i}^{t_f} \frac{\eta\,L}{T_{\rm rad}}\,dt
\end{equation}
\noindent If the emitted radiation is (on average) colder than the gas, each unit of energy carried away increases entropy by roughly \(\eta/T_{\rm rad}\), while extracting energy from the gas decreases its entropy by (roughly) \(1/T_{\rm gas}\) per unit energy (plus the \(P\,dV\) contribution). Since typically \(T_{\rm rad}\ll T_{\rm gas}\) for optically thick objects, the radiation term dominates and $\Delta S_{\rm tot}>0$.
\vskip0.15truecm
\noindent {\bf B) Limit of extreme contraction: Schwarzschild radius and Bekenstein - Hawking entropy}

\begin{figure}[h]
\center{
\includegraphics[width=9cm]{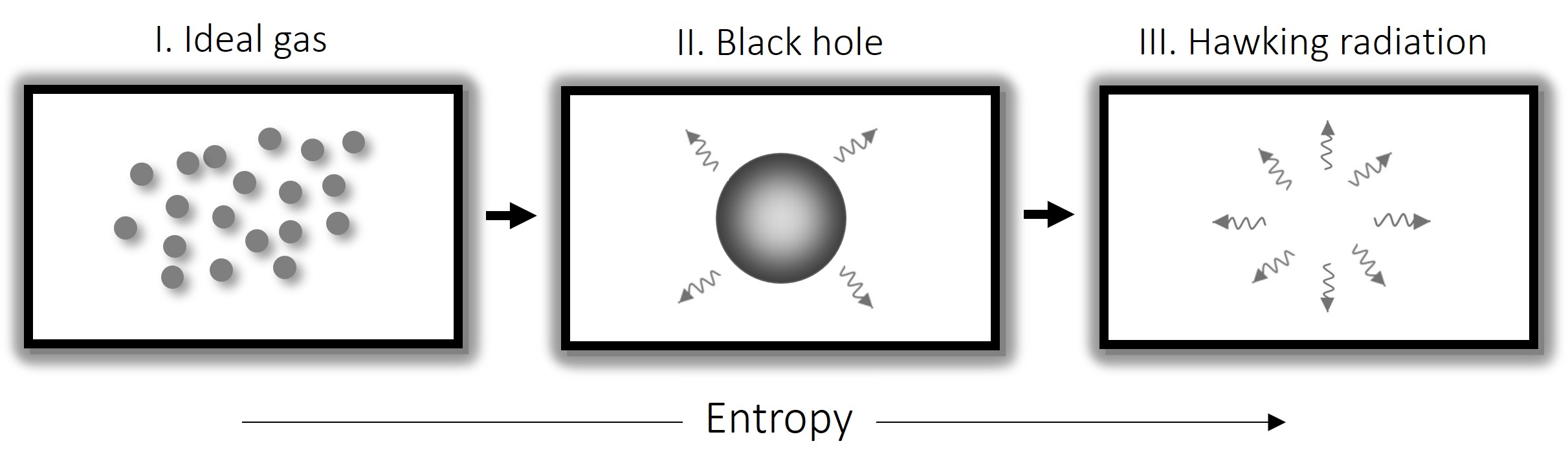}
\caption {Free contraction of a very large mass of ideal gas.}
\label{Fig5}}
\end{figure}

\noindent If contraction proceeds unchecked, relativistic gravity becomes important. The characteristic radius where the naive Newtonian description breaks down is the Schwarzschild radius,
\begin{equation}\label{WG14}
R_s = \frac{2GM}{c^2}
\end{equation}
\noindent  In 1973, theoretical physicist Jacob Bekenstein proposed that black holes have entropy by giving an approximate formula to calculate it \cite{bekenstein, carlip}. Shortly after, Stephen Hawking derived the exact expression for a black hole's entropy. In gravitational physics, the concept of entropy naturally extends to black holes. As first proposed by Jacob Bekenstein and later derived from quantum field theory in curved spacetime by Stephen Hawking, a black hole of mass $M$ and horizon area $A = 4\pi R_s^2$, has an entropy given by (for an intuitive, pedagogical derivation, see \cite{pinochet1, sonnino}):
\begin{equation}\label{eq:S_BH}
S_{\rm BH} = \frac{k_B c^3 A}{4G\hbar} = \frac{4\pi k_B G M^2}{\hbar c}
\end{equation}
\noindent This shows that a black hole’s entropy is proportional to its horizon area, highlighting that gravitational systems can harbor immense amounts of entropy and confirming that the second law of thermodynamics remains robust even in strongly gravitating regimes. Entropy~(\ref{eq:S_BH}) is enormous compared with the thermodynamic entropy of ordinary matter with the same mass. For instance, for a solar mass:
\begin{equation}\label{WG15}
\frac{S_{\rm BH}}{k_B} \approx \frac{4\pi G M_\odot^2}{\hbar c} \sim 10^{77}
\end{equation}
\noindent By contrast, the thermodynamic entropy of \(M_\odot\) worth of hydrogen gas at stellar conditions is
\begin{equation}\label{WG16}
S_{\rm gas} \sim N k_B \times O(10) \sim \frac{M_\odot}{m_p} k_B \times O(10) \sim 10^{58}\,k_B
\end{equation}
\noindent so
\begin{equation}\label{WG17}
S_{\rm BH} \gg S_{\rm gas} \quad\text{(roughly by } 10^{19}\text{ orders of magnitude)}
\end{equation}
\noindent Thus, gravitational collapse into a black hole \emph{increases} entropy enormously. The formation of a horizon hides information from an external observer, and (within semiclassical gravity) the horizon area encodes a vastly larger number of microstates. We are now in a position to state the generalized second law (GSL) for the black-holes: \textit{The sum of ordinary entropy outside black holes plus the Bekenstein-Hawking entropy associated with horizons never decreases} \cite{ruffini,wheeler}:
\begin{equation}\label{WG18}
\Delta\Big(S_{\rm outside} + \frac{k_B c^3 A}{4G\hbar}\Big) \ge 0
\end{equation}
\noindent This is the gravitational extension of the second law: when matter with entropy \(S_{\rm matter}\) falls into a black hole, the horizon area increases by an amount that (semiclassically) more than compensates for the lost external entropy. It is important to mention that Jacob Bekenstein and Stephen Hawking associated the black-hole entropy with the area of the event horizon. Considering that, according to the laws of thermodynamics, an object that has entropy must have a temperature, Hawking theoretically demonstrated that black holes have an absolute temperature that depends inversely on their mass according to the law \cite{sonnino}, \cite{hawking1, hawking2, pinochet2}:
\begin{equation}\label{WG19}
T_{H} = \frac{\hbar c^{3}}{8\pi kGM_{BH}}
\end{equation}
\noindent This expression reveals that black holes are not so black, and that they can be very luminous if $M_{BH}$ is small enough. We can calculate an upper bound for $T_{H}$ by considering a stellar black hole again. By plugging $M_{BH} \sim 10^{30}kg$ into Eq.~(\ref{WG19}) we get $T_{H}\sim 10^{-8} K$. This temperature is extremely small and undetectable by astronomical observations. As is known, every hot body emits thermal radiation, i.e., radiant energy, and black holes are no exception. This is the so-called Hawking radiation. But, by mass-energy equivalence, $E=mc^{2}$, this radiation has a mass equivalent. Thus, as the black hole emits Hawking radiation, it loses mass until it evaporates. Since $T_{H}$ depends inversely on $M_{BH}$, the emission rate increases over time. In theory, this process leads to the complete evaporation of the black hole with a characteristic time referred to as the evaporation time \cite{pinochet1, frolov}: 
\begin{equation}\label{WG20}
t_{ev} \approx 10^{3} \frac{G^{2}M_{BH}^{3}}{c^{4} \hbar}
\end{equation}
\noindent For $M_{BH} \sim 10^{30}kg$, the lower bound for the evaporation time is of the order of $t_{ev} \sim 10^{73}s$, i.e., 56 orders of magnitude larger than the age of the universe, which is of the order of $10^{17}s$. The above ideas allow us to answer the initial question: How does gas evolve in free contraction if the mass is very large? In general terms, and without going into the complex astrophysical processes governing the evolution of large masses of gas under the action of their own gravity, the sequence illustrated in Figure~\ref{Fig5} is as follows: 
\begin{itemize}
\item[$(I)$] {the gas mass contracts and heats, emitting thermal radiation;} 
\item [$(II)$]{A black hole is formed, which we can assume initially absorbs everything contained in the vessel, including the radiation emitted during phase I;}
\item [$(III)$]{The black hole evaporates completely, and inside the container, there is only Hawking radiation.}
\end{itemize}
\noindent These three stages are accompanied by an increase in entropy. Indeed, from the ideas developed in the previous sections, we know that the evolution from stage I to stage II in Fig. 5 entails a reduction in the volume of the gas and an increase in the total entropy inside the vessel. Denoting with $S_{G}$ and $S_{BH}$  the entropy of the gas in stage I and the entropy of the black hole in stage II, respectively, we can conclude that:

\begin{equation}\label{WG21}
\Delta S _{I \rightarrow II} = S_{BH}-S_{G} >0
\end{equation}
\noindent The evolution from stage II to III also involves an increase in entropy. If $S_{HR}$ is the entropy of Hawking radiation after complete evaporation, after quite complex calculations (that we shall not reproduce here), it is possible to show that $S_{HR} \cong 4S_{BH}/3$ \cite{zurek}. So: 
\begin{equation}\label{WG22}
\Delta S _{II \rightarrow III} = S_{HR}-S_{BH} = \frac{4S_{BH}}{3}-S_{BH} = \frac{S_{BH}}{3}>0
\end{equation}
\noindent By combining the last two results, we finally get:
\begin{equation}\label{WG23}
\Delta S _{I \rightarrow II} + \Delta S _{II \rightarrow III} >0
\end{equation}
\noindent Demonstrating the validity of the second law of thermodynamics. It is important to note that the three stages leading to the complete evaporation of a black hole have never been observed. As previously mentioned, witnessing this phenomenon would require a timescale far exceeding the age of the universe. Therefore, the process depicted in Fig. 5 remains purely theoretical, even though strong evidence supports it and it is widely accepted by the scientific community. The most remarkable aspect of these findings is that, in the very long term, uniformity appears to prevail despite gravity's tendency to resist it. It is worth noting that a consistent thermodynamic description of gravitational contraction requires taking into account that black holes are not isolated systems \cite{sonnino}. A self-gravitating gas undergoing contraction continuously loses energy through radiation (e.g., photons or neutrinos), thereby establishing an unavoidable coupling between matter and radiation. Consequently, any energy flux $L$ leaving the system must be accompanied by an entropy flux $\eta L/T_{\rm rad}$, and the total entropy variation must include both the gas contribution and the entropy carried away by radiation:
\begin{equation}
\frac{dS_{\rm tot}}{dt} = \frac{dS_{\rm gas}}{dt} + \frac{dS_{\rm rad}}{dt} \ge 0.
\end{equation}
\noindent This balance is naturally interpreted within the framework of non-equilibrium thermodynamics for open systems, as formulated by Prigogine, where the total entropy change is decomposed as \cite{prigogine}
\begin{equation}
dS_{\rm tot} = dS_{\rm production} + dS_{\rm flux}, \qquad dS_{\rm production} \ge 0.
\end{equation}
\noindent In this context, the entropy carried away by radiation corresponds to the flux term,
\begin{equation}
\frac{dS_{\rm flux}}{dt} = \eta\frac{L}{T_{\rm rad}},
\end{equation}
\noindent while the gas contribution accounts for the internal entropy changes and irreversible production. This formulation makes clear that neglecting radiation would amount to treating a fundamentally open system as isolated. Within this unified picture, black hole thermodynamics can be viewed as a limiting case in which gravitational collapse leads to an extreme concentration of entropy at the horizon, while the generalized second law remains preserved through the combined contributions of matter, radiation, and the black hole itself.

\vskip0.15truecm
\noindent {\bf From Protostellar Collapse to Neutrino Cooling}

\noindent Before starting calculations, for easy reference, we recall the values of the relevant constants entering the calculations:

\begin{equation}\label{excd1}
\boxed{
\begin{aligned}
G &= 6.67430\times10^{-11}\ \mathrm{m^3\,kg^{-1}\,s^{-2}}\\
M_\odot &= 1.98847\times10^{30}\ \mathrm{kg}\\
R_\odot &= 6.957\times10^{8}\ \mathrm{m}\\
m_p &= 1.67262\times10^{-27}\ \mathrm{kg}\\
k_B &= 1.38065\times10^{-23}\ \mathrm{J\,K^{-1}}\\
h &= 6.62607\times10^{-34}\ \mathrm{J\,s}\\
1\,\mathrm{eV} &= 1.16045\times10^{4}\ \mathrm{K}
\end{aligned}
}
\end{equation}

\vskip0.15truecm
\noindent {\bf C) Protostellar contraction sequence}

\noindent We study a proto-star of mass \(M = 1\,M_\odot\) contracting from \(R_i = 3R_\odot\) to \(R_f = R_\odot\). Assume a uniform-density sphere with potential energy \cite{stahler, kippenhahn}:
\begin{equation}\label{excd2}
U = -\alpha \frac{G M^2}{R}, \qquad \alpha = \frac{3}{5}
\end{equation}
\noindent By the virial theorem $2K + U = 0$, so the total energy is $E = K + U = U/2$. The mean thermal temperature is approximately
\begin{equation}\label{excd3}
T(R) = \frac{\alpha G M m_p}{3 k_B R}
\end{equation}
\noindent Substituting numbers, we get:
\begin{equation}\label{excd3}
T_i = 1.54\times10^{6}\ \mathrm{K},\qquad T_f = 4.62\times10^{6}\ \mathrm{K}
\end{equation}
\noindent Let us now compute the gravitational energy release.
\begin{align}\label{excd4}
U_i &= -7.59\times10^{40}\ \mathrm{J},\qquad\qquad\quad\   U_f = -2.28\times10^{41}\ \mathrm{J}\\
E_i &= U_i/2 = -3.79\times10^{40}\ \mathrm{J},\qquad E_f = U_f/2 = -1.14\times10^{41}\ \mathrm{J}\nonumber
\end{align}
\noindent The energy radiated away reads:
\begin{equation}\label{excd5}
E_{\rm rad} = E_i - E_f = 7.59\times10^{40}\ \mathrm{J}
\end{equation}
\noindent If this occurs over \(t_{\rm KH} = 3\times10^6\ \mathrm{yr}\), the average luminosity is:
\begin{equation}\label{excd6}
\bar L = \frac{E_{\rm rad}}{t_{\rm KH}} \approx 8.0\times10^{26}\ \mathrm{W}
\end{equation}
\noindent of the order of the solar luminosity. The entropy of the gas can be computed by using the Sackur–Tetrode formula for \(N = M/m_p = 1.19\times10^{57}\) hydrogen atoms:
\begin{equation}\label{excd7}
S = N k_B\!\left[\ln\!\left(\frac{V}{N\lambda^3}\right)+\frac{5}{2}\right],
\qquad
\lambda = \frac{h}{\sqrt{2\pi m_p k_B T}}
\end{equation}
\noindent
\begin{align}\label{excd8}
V_i &= \frac{4\pi}{3}R_i^3 = 3.81\times10^{28}\ \mathrm{m^3},\qquad V_f = \frac{4\pi}{3}R_f^3 = 1.41\times10^{26}\ \mathrm{m^3}\\
\lambda_i &= 1.40\times10^{-12}\ \mathrm{m},\qquad\qquad\quad\ \  \lambda_f = 8.09\times10^{-13}\ \mathrm{m}\nonumber
\end{align}
\noindent Evaluating the change in entropy, we find:
\begin{equation}\label{excd9}
S_i = 3.08\times10^{35}\ \mathrm{J\,K^{-1}},\quad S_f = 2.81\times10^{35}\ \mathrm{J\,K^{-1}}\quad\Rightarrow\quad \Delta S_{\rm gas} = -2.70\times10^{34}\ \mathrm{J\,K^{-1}}
\end{equation}
\noindent Thus, the gas entropy decreases as the protostar contracts. The entropy radiation is also easily computed. Assuming the radiative temperature $T_{\rm rad} \approx 4000\ \mathrm{K}$, we get:
\begin{equation}\label{excd10}
S_{\rm rad} = \frac{E_{\rm rad}}{T_{\rm rad}}
= \frac{7.59\times10^{40}}{4.0\times10^{3}}
= 1.90\times10^{37}\ \mathrm{J\,K^{-1}}
\end{equation}
\noindent Finally, the total entropy change is:
\begin{equation}\label{excd11}
\Delta S_{\rm tot} = \Delta S_{\rm gas} + S_{\rm rad}
= (-2.7\times10^{34}) + 1.9\times10^{37}
\approx +1.9\times10^{37}\ \mathrm{J\,K^{-1}}
\end{equation}
\noindent Hence, although the gas becomes more ordered, the emitted radiation produces an entropy increase about \(10^3\) times larger, preserving the second law.

\vskip0.15truecm
\noindent {\bf D) Core collapse and neutrino cooling}

\noindent In a core-collapse supernova, a degenerate iron core of $M_{\rm core} \approx 1.4 M_\odot$
collapses to a neutron star of radius \(R_{\rm NS} \approx 12\ \mathrm{km}\). The value of the binding energy is \cite{suzuki, reed}:
\begin{equation}\label{exn1}
|U| = \alpha \frac{G M^2}{R} 
= 0.6 \times \frac{(6.674\times10^{-11})(2.78\times10^{30})^2}{1.2\times10^{4}}
\approx 2.59\times10^{46}\ \mathrm{J}
\end{equation}
\noindent Half of this is radiated as neutrinos:
\begin{equation}\label{exn2}
E_\nu \approx 1.3\times10^{46}\ \mathrm{J}
\end{equation}
\noindent Typical neutrino energies are a few MeV. For \(T_\nu = 4\,\mathrm{MeV} = 4.64\times10^{10}\ \mathrm{K}\), the neutrino's entropy is 
\begin{equation}\label{exn3}
S_\nu = \frac{E_\nu}{T_\nu}
= \frac{1.3\times10^{46}}{4.64\times10^{10}}
\approx 2.8\times10^{35}\ \mathrm{J\,K^{-1}}
\end{equation}
\noindent The baryonic matter inside the collapsing core loses entropy as it becomes highly degenerate, but this reduction ($\lesssim 10^{33}\ \mathrm{J\,K^{-1}}$) is negligible compared with the neutrino entropy. Thus, the matter entropy change reads:
\begin{equation}\label{exn4}
\Delta S_{\rm tot} \approx S_\nu \approx 10^{35}\text{-}10^{36}\ \mathrm{J\,K^{-1}}
\end{equation}
\noindent The neutrino burst dominates the entropy budget and guarantees \(\Delta S_{\rm tot} > 0\). To summarize:
\begin{itemize}
\item [$-$] Gravitational contraction decreases the entropy of matter (gas becomes denser and more ordered).
\item [$-$] Energy conservation requires that gravitational binding energy be released as radiation (photons or neutrinos).
\item [$-$] The entropy carried by emitted radiation far exceeds the entropy lost by matter.
\item [$-$] Thus, even during gravitational collapse, the \emph{total} entropy of the system + environment increases, in full agreement with the second law.
\item [$-$] In the ultimate limit of black-hole formation, the Bekenstein-Hawking entropy,
\begin{equation}\label{exn5}
S_{\rm BH} = \frac{k_B c^3 A}{4 G \hbar} = 1.05\times10^{77}\left(\frac{M}{M_\odot}\right)^2 k_B
\end{equation}
\noindent dominates, ensuring that the final state possesses the largest possible entropy.
\end{itemize}

\vskip0.15truecm 
\noindent {\bf Limitations and concluding remarks}

\noindent
\begin{itemize}
\item [$-$] {Quasi-static assumption:} our derivation of \(\dot S_{\rm gas}=(dE+PdV)/T\) assumes the gas remains near local thermodynamic equilibrium. For violent, highly non-equilibrium collapse (shock formation, fragmentation), one must treat irreversible processes explicitly; however, irreversible processes \emph{increase} entropy, making the second law easier to satisfy.
\item [$-$] {Opacity and photon trapping:} if the system is optically thick, photons are temporarily trapped and the effective \(T_{\rm rad}\) at the photosphere can be much lower than the interior temperature; trapped radiation contributes to internal energy until diffusion/transport allows escape. Neutrinos and convective transport can dominate cooling in some regimes (e.g.\ core-collapse supernovae). The formalism above still applies: \(L\) is the net power leaving the system and \(T_{\rm rad}\) is the effective temperature of escaping carriers.
\item [$-$] {Nonthermal emission:} if radiation is nonthermal (line emission, bremsstrahlung with hard spectrum), the factor \(\eta\) in \(\dot S_{\rm rad}=\eta L/T_{\rm rad}\) should be evaluated suitably; in many practical cases the entropy per unit energy for escaping radiation is larger than \(1/T_{\rm gas}\) because photons are emitted at lower mean energy than local thermal energy per degree of freedom.
\item [$-$] {Gravitational waves:} energy can also be radiated as gravitational waves. Their entropy accounting is subtler; however, gravitational waves typically carry little entropy compared with photons/neutrinos for the same energy, so they do not typically undermine the global entropy increase.
\item [$-$]{Black hole formation:} once an event horizon forms, the microphysical description of entropy becomes semiclassical; whether horizon entropy counts true microstates or accounts for entanglement entropy remains an active area of research. The generalized second law is the operational statement ensuring second law consistency, including horizons.
\end{itemize}
\noindent Furthermore,
\begin{itemize}
\item [$-$] Gravitational contraction tends to decrease the thermodynamic entropy of the \emph{matter} (gas) because volume decreases and particles become more spatially ordered.
\item [$-$] Conservation of energy forces contracting matter to \emph{release} energy; this energy is typically emitted as radiation (photons, neutrinos), and the emitted radiation carries a large amount of entropy per unit energy (especially if the radiation is emitted at temperatures much lower than the interior temperature).
\item [$-$] When one accounts for the radiative entropy flux, the total entropy of the closed system (matter $+$ radiation) increases:
\begin{equation}\label{CR1}
\dot S_{\rm tot} = \dot S_{\rm gas} + \dot S_{\rm rad} \ge 0
\end{equation}
\noindent explicitly given approximately by \eqref{eq:Sdot_key}.
\item [$-$] In the extreme limit of black hole formation, entropy becomes dominated by the horizon (Bekenstein--Hawking) entropy, which vastly exceeds the matter entropy; the generalized second law ensures the total never decreases.
\end{itemize}

\section{Conclusions}

We have demonstrated that, through simple calculations and intuitive reasoning, it is possible to uncover fundamental insights into the relationship between entropy and gravity. In particular, we have shown that gravity does not violate the second law of thermodynamics; rather, gravitational systems adhere to and often exemplify this law in unique ways. We hope that this work helps bridge, at least in part, the existing gap in the literature on this topic and provides students with a deeper understanding of thermodynamics. For easy reference, we report the practical formula useful for students:
\begin{equation}\label{c1}
\boxed{\quad \dot S_{\rm tot} \;=\; L\Big(\frac{\eta}{T_{\rm rad}} - \frac{1}{T_{\rm gas}}\Big) + \frac{P\,\dot V}{T_{\rm gas}} \quad}
\end{equation}
\noindent with \(\eta\approx 4/3\) for blackbody emission. This expression displays clearly where the missing entropy sits: in the term $L\,\eta/T_{\rm rad}$ -  the entropy carried away by the emitted energy. We conclude with the following observation. We have seen that, by combining the rates of entropy change of the gas and the radiation, the total entropy production remains non-negative under standard physical assumptions (namely, quasi-static contraction and radiation escaping from the system). The compensating entropy is carried away by the emitted energy in the form of photons, neutrinos, gravitational waves, and so forth. 


\appendix
\section {A Simple Derivation of the Sackur-Tetrode Entropy}

In this appendix, we provide a compact but careful derivation of the Sackur-Tetrode entropy for a monatomic ideal gas starting from the canonical partition function. The Stirling approximation will be kept explicitly, and the thermal wavelength $\lambda$ will be shown clearly. Let us begin by considering the thermal wavelength of a single particle. We define the (thermal) de Broglie wavelength:
\begin{equation}\label{a1}
\lambda(T) \equiv \frac{h}{\sqrt{2\pi m k_B T}}.
\end{equation}
\noindent This is convenient because the single-particle translational partition function can be written as:
\begin{equation}\label{a2}
z_1 = \frac{V}{\lambda^3}.
\end{equation}
\noindent Let us consider the canonical partition function for $N$ indistinguishable particles. Assuming classical behavior, with indistinguishability corrected by \(1/N!\):
\begin{equation}\label{a3}
Z(T,V,N) = \frac{1}{N!} \left( \frac{V}{\lambda^3} \right)^N
\end{equation}
\noindent The Helmholtz free energy reads:
\begin{equation}\label{a4}
F = -k_B T \ln Z = -k_B T \big[ N \ln V - 3N \ln \lambda - \ln N! \big]
\end{equation}
\noindent The expression for the entropy $S$ at fixed $V$ is given by
\begin{equation}\label{a5}
S = -\left( \frac{\partial F}{\partial T} \right)_V.
\end{equation}
\noindent  We note that $\ln N!$ is $T$-independent, while $\ln \lambda$ depends on $T$. The computation of $\partial \ln \lambda / \partial T$ yields:
\begin{equation}\label{a6}
\ln \lambda = \ln h - \frac{1}{2}\ln(2\pi m k_B T)\quad \Rightarrow \quad \frac{\partial \ln \lambda}{\partial T} = -\frac{1}{2T}
\end{equation}
\noindent Differentiating $F$ we get:
\begin{equation}\label{a7}
\frac{\partial F}{\partial T}= -k_B \big[N\ln V - 3N\ln\lambda - \ln N!\big]- k_B T \big[-3N\, \frac{\partial \ln\lambda}{\partial T}\big]
\end{equation}
\noindent Plugging in \(\frac{\partial \ln \lambda}{\partial T} = -\frac{1}{2T}\) gives
\begin{equation}\label{a8}
\frac{\partial F}{\partial T}
= -k_B N\ln V + 3k_B N\ln\lambda + \frac{3}{2}k_B N - k_B \ln N!
\end{equation}
\noindent Hence,
\begin{equation}\label{a9}
S = -\frac{\partial F}{\partial T}
= k_B \big[N\ln V - 3N\ln\lambda - \tfrac{3}{2}N + \ln N!\big]
\end{equation}
\noindent We shall rearrange this expression by applying Stirling’s approximation. By using Stirling’s formula for large $N$, we find
\begin{equation}\label{a10}
\ln N! = N\ln N - N + \frac{1}{2}\ln(2\pi N) + \mathcal{O}(1/N)
\end{equation}
\noindent Substitute this into the expression for \(S\), we have:
\begin{equation}\label{a11}
\frac{S}{k_B}
= N\ln V - 3N\ln\lambda - \frac{3}{2}N + \big(N\ln N - N + \tfrac{1}{2}\ln(2\pi N)\big).
\end{equation}
\noindent Simplifying:
\begin{equation}\label{a12}
\frac{S}{k_B}= N\ln\frac{V}{N} - 3N\ln\lambda + \frac{5}{2}N + \frac{1}{2}\ln(2\pi N)
\end{equation}
\noindent It is customary to drop the sub-leading \(\frac{1}{2}\ln(2\pi N)\) term for macroscopic \(N\) (it is negligible compared to \(N\)). Doing so yields the usual Sackur--Tetrode form. We neglect the finite-$N$ correction:
\begin{equation}\label{a13}
\boxed{S = N k_B \left[ \ln\!\left(\frac{V}{N\lambda^3}\right) + \frac{5}{2} \right]}
\end{equation}
\noindent This is the standard Sackur-Tetrode entropy for a monatomic ideal gas (canonical ensemble). We note the following points:
\begin{itemize}
    \item [$-$] The \(\ln(V/N\lambda^3)\) term counts translational quantum cells of volume \(\lambda^3\) per particle and resolves the Gibbs paradox by including the \(1/N!\) correction.
    \item [$-$] The constant \(5/2\) arises from combining the internal energy dependence (degrees of freedom) with the Stirling linear term.
\end{itemize}

\noindent If we keep the finite-\(N\) Stirling correction, the more precise form is:
\begin{equation}\label{a14}
S = N k_B\left[\ln\!\left(\frac{V}{N\lambda^3}\right) + \frac{5}{2}\right] + \frac{1}{2}k_B\ln(2\pi N) + \mathcal{O}\!\left(\frac{1}{N}\right)
\end{equation}
\noindent It is instructive now to rewrite the entropy expression in \textit{microcanonical form} by replacing $k_B T$ with $2U/3N$. Using \(U = \tfrac{3}{2} N k_B T\), so \(k_B T = \tfrac{2U}{3N}\). Substitute into the canonical form to get the microcanonical variant. Since \(\lambda^3 = h^3 / (2\pi m k_B T)^{3/2}\), we find after simple algebra:
\begin{equation}\label{a15}
\boxed{
S = N k_B \left[\ln\!\left(\frac{V}{N}\left(\frac{4\pi m U}{3N h^2}\right)^{3/2}\right) + \frac{5}{2}\right]
}
\end{equation}
\noindent where we have taken into account that one step that produces the \(4\pi\) is \((2\pi)^{3/2}(2)^{3/2}=(4\pi)^{3/2}\), so the factors combine neatly. We also note that because \(\lambda\) depends only on \(T\), for an isothermal free expansion (\(T\) constant) the change of Sackur-Tetrode entropy between \(V_i\) and \(V_f\) reduces to:
\begin{equation}\label{a16}
\Delta S = N k_B \ln\frac{V_f}{V_i}
\end{equation}
\noindent in exact agreement with the thermodynamic/macroscopic result.
\vskip0.15truecm
\noindent {\bf Physical meaning of the obtained results}

\noindent It is instructive and beneficial for students to outline the physical meaning of the results obtained briefly.

\begin{itemize}
    \item [$-$] The appearance of Planck’s constant \(h\) is a counting device ensuring the correct quantum size of phase-space cells; it resolves the Gibbs paradox.
    \item [$-$] The Sackur-Tetrode formula is valid in the classical (nondegenerate) limit, when the thermal wavelength $\lambda$ is much smaller than the interparticle spacing.
    \item [$-$] Finite-$N$ corrections ($\tfrac{1}{2}\ln(2\pi N)$ term) are negligible for macroscopic systems but can matter for small systems.
\end{itemize}

\section{Energy Balance, the Virial Theorem, and the Origin of the Factor $1/2$}

Let us unpack and justify carefully where the factor of $1/2$ comes from, and when 
\begin{equation}\label{b1}
dK = -\frac{dU}{2}
\end{equation}
\noindent rather than 
\begin{equation}\label{b2}
dK = -dU.
\end{equation}
\noindent For any isolated system (no external work or radiation losses),
\begin{equation}\label{b3}
dE = d(K + U) = 0 \quad \Rightarrow \quad dK = -dU
\end{equation}
\noindent This is a statement of instantaneous energy conservation: if the total energy \(E\) is conserved, any increase in potential energy (less negative \(U\)) must be matched by a decrease in kinetic energy, and vice versa.  Hence, for instantaneous changes, the correct differential relation is $dK = -dU$. However, the \emph{virial theorem} does not refer to instantaneous changes but to time averages (or equivalently, to equilibrium configurations in quasi-static evolution). For a bound, virialized system with potential \(U(r) \propto r^{-1}\) (as in Newtonian gravity), the scalar virial theorem states:
\begin{equation}\label{b4}
2 \langle K \rangle + \langle U \rangle = 0
\end{equation}
\noindent Here, the angle brackets \(\langle \cdot \rangle\) denote time averages over many dynamical times. This implies:
\begin{equation}\label{b5}
\langle K \rangle = -\frac{1}{2} \langle U \rangle.
\end{equation}
\noindent Therefore, the total energy is:
\begin{equation}\label{b6}
E = \langle K \rangle + \langle U \rangle = \frac{1}{2}\langle U \rangle = -\langle K \rangle
\end{equation}
\noindent When one speaks of a \emph{quasi-static contraction}, for example, a star slowly shrinking under gravity while staying close to virial equilibrium, we assume that at each moment, the virial relation $2K + U = 0$ approximately holds. Differentiating this virial relation gives:
\begin{equation}\label{b7}
2\,dK + dU = 0 \quad \Rightarrow \quad dK = -\frac{1}{2}\,dU
\end{equation}
\noindent Thus, when the system passes from one virialized equilibrium to another, only half of the released gravitational potential energy goes into increasing the kinetic (thermal) energy; the other half must be radiated away to maintain equilibrium. This is the logic used in stellar physics to derive the Kelvin-Helmholtz timescale. For an ideal gas, the mean kinetic energy per particle is given by the equipartition theorem:
\begin{equation}\label{b8}
\langle \epsilon_k \rangle = \frac{3}{2} k_B T.
\end{equation}
\noindent Thus, for \(N\) particles,
\begin{equation}\label{b9}
K = \frac{3}{2} N k_B T
\end{equation}
\noindent This relation holds for a classical, non-degenerate gas in local thermodynamic equilibrium. Therefore, identifying \(K\) with the thermal energy is appropriate for protostellar and many astrophysical contexts. We can summarize the distinctions between the instantaneous and virial-averaged relations as follows:
\begin{center}
\begin{tabular}{@{}lll@{}}
\toprule
\textbf{Aspect} & \textbf{Description} & \textbf{Relation} \\
\midrule
Instantaneous energy conservation & Holds always & \(dK = -dU\) \\
Virial equilibrium (time-averaged) & For stable, bound systems & \(2K + U = 0\) \\
Quasi-static contraction sequence & Evolution through successive equilibria & \(dK = -\tfrac{1}{2} dU\) \\
Thermal interpretation & Defines temperature & \(K = \tfrac{3}{2} N k_B T\) \\
\bottomrule
\end{tabular}
\end{center}
\noindent Hence, the relation \(dK = -dU/2\) is not in conflict with energy conservation - it reflects the \emph{average} or \emph{quasi-static} limit, where the system maintains near-virial equilibrium and radiates away half of the gravitational energy released during contraction. As a star contracts:
\begin{enumerate}
    \item Gravitational potential energy \(U\) becomes more negative (\(dU < 0\));
    \item Half of \(-dU\) goes into increasing internal thermal energy (\(dK > 0\));
    \item The other half is radiated away:
    \begin{equation}\label{b10}
        dE_{\text{rad}} = -dK - dU = -\frac{dU}{2}
    \end{equation}
\end{enumerate}
\noindent Thus, the luminosity of a contracting star can be estimated from:
\begin{equation}\label{b11}
L \approx -\frac{dE}{dt} = -\frac{1}{2}\frac{dU}{dt}
\end{equation}
\noindent The system radiates away roughly half of the gravitational energy released during contraction - consistent with both the virial theorem and the second law of thermodynamics. In the next susection, we shall see that, with explicit timescale separation, in a protostar we have:
\begin{equation}\label{b12}
t_{\text{coll}}\ll t_{\text{dyn}}\ll t_{\text{KH}}
\end{equation}
\noindent That is, microscopic equilibration (collisions) and dynamical readjustments occur orders of magnitude faster than the secular contraction driven by radiative losses. Hence, the protostar evolves through a sequence of quasi-static equilibria - meaning that at each stage, the virial theorem $2K+U\approx 0$ holds to excellent approximation. Thus, using the virial relation and its differential form $dK=-1/2dU$ is fully justified for the system’s slow, thermally regulated contraction phase. This is exactly how stellar structure textbooks treat pre-main-sequence contraction (see \cite{stahler, kippenhahn}).

\subsection{Timescale Estimates Justifying the Quasi-static Virial Approximation}

\noindent In this section, we present the order-of-magnitude timescale estimates for a protostellar contraction example used in the text:
\begin{equation}\label{bb1}
M = 1\,M_\odot,\qquad R_i = 3R_\odot,\qquad R_f = 1R_\odot
\end{equation}
\noindent where \(R_\odot=6.957\times10^8\ \mathrm{m}\) and \(M_\odot=1.98847\times10^{30}\ \mathrm{kg}\). We adopt the uniform-sphere approximation for simplicity; the results are representative and sufficient to verify the separation of timescales. Furthermore, in our estimates, we shall use the following numerical values:
\noindent 
\(G=6.67430\times10^{-11}\ \mathrm{m^3\,kg^{-1}\,s^{-2}}\),
\(m_p=1.6726\times10^{-27}\ \mathrm{kg}\),
\(k_B=1.38065\times10^{-23}\ \mathrm{J\,K^{-1}}\),
\(L_\odot=3.83\times10^{26}\ \mathrm{W}\).

\noindent The values shown above keep two–three significant figures; estimates are intentionally order-of-magnitude.
\vskip0.15truecm
\noindent{\bf Characteristic timescales and characteristic temperature}

\noindent Let us estimate the \textit{dynamical (free-fall/crossing) time}. For a self-gravitating object of mass \(M\) and radius \(R\) the dynamical time (order of magnitude) is
\begin{equation}\label{bb2}
t_{\rm dyn} \sim \sqrt{\frac{R^3}{G M}}.
\end{equation}
\noindent Evaluating numerically gives
\begin{equation}\label{bb2}
t_{\rm dyn}(R_i=3R_\odot) \approx 8.3\times10^3\ \mathrm{s}\ \approx 2.3\ \mathrm{hr},\qquad
t_{\rm dyn}(R_f=1R_\odot)   \approx 1.59\times10^3\ \mathrm{s}\ \approx 26.5\ \mathrm{min}
\end{equation}
\noindent We now estimate the \textit{virial (characteristic) temperature and sound speed}. Using the virial estimate (with \(\alpha=3/5\) for a uniform sphere):
\begin{equation}\label{bb3}
T(R)=\frac{\alpha G M m_p}{3 k_B R},
\end{equation}
\noindent with \(m_p\) denoting the proton mass and \(k_B\) Boltzmann constant, respectively we obtain
\begin{equation}\label{bb4}
T_i\equiv T(R_i) \approx 1.54\times10^{6}\ \mathrm{K},\quad T_f\equiv T(R_f) \approx 4.62\times10^{6}\ \mathrm{K}
\end{equation}
\noindent The isothermal sound speed estimate \(c_s\simeq\sqrt{k_B T/m_p}\) yields:
\begin{align}\label{bb5}
c_{s,i} &\approx 1.13\times10^5\ \mathrm{m\,s^{-1}}, & 
t_{\rm sound,i}\equiv \frac{R_i}{c_{s,i}} &\approx 1.85\times10^4\ \mathrm{s}\ (\approx 5.14\ \mathrm{hr}),\\
c_{s,f} &\approx 1.95\times10^5\ \mathrm{m\,s^{-1}}, & 
t_{\rm sound,f}\equiv \frac{R_f}{c_{s,f}} &\approx 3.56\times10^3\ \mathrm{s}\ (\approx 59\ \mathrm{min})\nonumber
\end{align}
\noindent The \textit{Kelvin--Helmholtz} (KH) \textit{thermal timescale} is the time to radiate the gravitational binding energy at luminosity \(L\). Using \(E_{\rm grav}\sim G M^2/R\) and the present solar luminosity \(L_\odot\approx 3.83\times10^{26}\ \mathrm{W}\) as a reference,
\begin{equation}\label{bb6}
t_{\rm KH}\sim\frac{G M^2}{R L}.
\end{equation}
\noindent Numerically (using \(L=L_\odot\)) we get:
\begin{equation}\label{bb7}
t_{\rm KH}(R_i=3R_\odot) \approx 3.30\times10^{14}\ \mathrm{s}\ \approx 1.05\times10^{7}\ \mathrm{yr},\ \ t_{\rm KH}(R_f=1R_\odot)   \approx 9.91\times10^{14}\ \mathrm{s}\ \approx 3.14\times10^{7}\ \mathrm{yr}
\end{equation}
\noindent Let us now estimate the \textit{collision/relaxation time and local thermalization}. Using a rough interparticle cross section \(\sigma\sim 10^{-19}\ \mathrm{m^2}\) (order-of-magnitude for neutral hydrogen interactions in a dense gas) and the mean number density \(n\), one can estimate a kinetic collision time
\begin{equation}\label{bb8}
t_{\rm coll}\sim\frac{1}{n\sigma v_{\rm th}}
\end{equation}
\noindent where \(v_{\rm th}\sim c_s\). For the uniform-sphere densities implied by \(M=1M_\odot\),
\begin{equation}\label{bb9}
\rho_i \sim \frac{M}{\tfrac{4}{3}\pi R_i^3}\approx 52\ \mathrm{kg\,m^{-3}},
\qquad n_i=\rho_i/m_p \sim 3.1\times10^{28}\ \mathrm{m^{-3}}
\end{equation}
\noindent giving (with the numbers above)
\begin{equation}\label{bb10}
\lambda_{\rm mfp}\sim\frac{1}{n_i\sigma}\sim 3\times10^{-10}\ \mathrm{m},\qquad
t_{\rm coll}\sim \frac{\lambda_{\rm mfp}}{v_{\rm th}}\sim 3\times10^{-15}\ \mathrm{s}
\end{equation}
\noindent Even allowing orders-of-magnitude uncertainty in \(\sigma\), the collision time is vastly shorter than dynamical and sound-crossing times, so the gas establishes (local) thermodynamic equilibrium on microscopic timescales. We are now in a position to separate the timescales and assess their implications. Collecting the characteristic times (representative values):
\begin{equation}\label{bb11}
t_{\rm coll}\ll t_{\rm dyn}\sim\mathrm{hours}\ll t_{\rm sound}\sim\mathrm{hours}\ll t_{\rm KH}\sim 10^{7}\text{--}10^{8}\ \mathrm{yr}
\end{equation}
\noindent Since
\begin{equation}\label{bb12}
t_{\rm dyn},\,t_{\rm sound} \ll t_{\rm KH},
\end{equation}
\noindent the protostar evolves on a timescale (Kelvin--Helmholtz time) that is many orders of magnitude longer than the internal dynamical and local equilibration times. Consequently:
\begin{itemize}
  \item The gas has time to (re-)establish a near-Maxwellian velocity distribution and local thermodynamic equilibrium during the contraction. In this appendix, we have explicitly stated that local thermodynamic equilibrium (LTE) is maintained because $t_{\text{coll}}\ll t_{\text{dyn}}$. Therefore, the velocity distribution of particles remains Maxwellian, so that the equipartition relation $K=3/2Nk_BT$ remains valid locally, even in a gravitational potential. The gravitational interaction couples regions, but does not modify the microscopic equipartition relation for a collisional gas in LTE - a well-established result in kinetic theory and astrophysical fluid dynamics. Thus, identifying kinetic energy with thermal energy is physically sound in the quasi-static, collisional regime.
  \item Pressure forces and internal motions (sound waves) can react on timescales short compared with the contraction, keeping the configuration near hydrostatic/virial balance.
  \item Hence, the quasi-static assumption (evolution through a sequence of near-virial equilibria) is well justified for protostellar Kelvin-Helmholtz contraction; under these conditions, the virial relation \(2K+U\approx 0\) and the differential form \(dK=-\tfrac{1}{2}dU\) are appropriate approximations.
\end{itemize}
\noindent If the collapse is \emph{violent} (e.g., rapid free-fall, strong shocks, fragmentation), then \(t_{\rm contract}\sim t_{\rm dyn}\) and the above quasi-static/virial argument is not valid - a full dynamical treatment (hydrodynamics / radiative transfer / possibly $N$-body) is required. We may object that the total potential energy is not linear in $N$; hence, thermodynamics cannot be defined for self-gravitating systems. This is conceptually correct in the strict sense (gravitational systems are non-extensive), but irrelevant to the present local argument. We are not constructing a thermodynamic limit - you are using hydrostatic equilibrium and virial relations for a finite, self-gravitating, collisional gas (a star), where the total potential energy $U-\alpha(GM^2)/R$ is by definition proportional to $M^2$, not $N$, as used in standard stellar structure theory. This scaling is already built into the virial theorem and the Kelvin–Helmholtz argument used universally in astrophysics. Hence, there is no inconsistency - we are not assuming extensivity, only applying the usual scaling laws valid for self-gravitating fluids.



\begin{thebibliography}{alpha}

\bibitem{resnick} R. Resnick, D. Halliday, K.S. Krane, Physics, 4th ed., John Wiley and Sons, New York, 1992.
\bibitem{tipler} P.A. Tipler, Physics for Scientists and Engineers, W. H. Freeman and Company, New York, 2004.
\bibitem{young} H.D. Young, R.A. Freedman, Sears and Zemansky’s University physics with modern physics, 14th ed., Pearson, New Jersey, 2016.
\bibitem{wilson} J.D. Wilson, A.J. Buffa, B. Lou, College Physics, 6th ed., Pearson, New Jersey, 2007.
\bibitem{walker} J. Walker, Fundamentals of physics, 9th ed., John Wiley and Sons, Hoboken, 2011.
\bibitem{bauer} W. Bauer, G. Westfall, University physics with modern physics, 1st ed., McGraw-Hill, New York, 2011.
\bibitem{reif} F. Reif, Fundamentals of Statistical and Thermal Physics, Waveland Press, Illinois, 2009.
\bibitem{schroeder} D.V. Schroeder, An Introduction to Thermal Physics, Oxford University Press, Oxford, 2021.
\bibitem{kittel} C. Kittel, W.D. Knight, M.A. Ruderman, A.C. Helmholz, B.J. Moyer, Mechanics, 2nd ed., McGraw-Hill, New York, 1973.
\bibitem{maoz} D. Maoz, Astrophysics in a Nutshell, Princeton University Press, Princeton, 2016.
\bibitem{bekenstein} J.D. Bekenstein, Black Holes and Entropy, Physical Review D 7 (1973) 2333–2346.
\bibitem{carlip} S. Carlip, Black hole thermodynamics, Int. J. Mod. Phys. D 23 (2014) 1430023. https://doi.org/10.1142/S0218271814300237.
\bibitem{pinochet1} J. Pinochet, Black holes ain’t so black: An introduction to the great discoveries of Stephen Hawking, Phys. Educ. 54 (2019) 035014.
\bibitem{sonnino} G. Sonnino, Prigogine’s Second Law and Determination of the EUP and GUP Parameters in Small Black Hole Thermodynamics, Universe 10 (2024) 390. https://doi.org/10.3390/universe10100390.
\bibitem{ruffini} D. Christodoulou and R. Ruﬃni, Reversible Transformations of a Charged Black Hole, Phys. Rev. D, 4, (1971).
\bibitem{wheeler} C.W. Misner, K.S. Thorne, and J.A. Wheeler, Gravitation, W. H. Freeman and Company, San Francisco (1973).
\bibitem{hawking1}  S.W. Hawking, Black Hole explosions?, Nature 248 (1974) 30–31.
\bibitem{hawking2} S.W. Hawking, Particle creation by black holes, Communications in Mathematical Physics 43 (1975) 199–220.
\bibitem{pinochet2} J. Pinochet, Hawking for everyone: commemorating half a century of an unfinished scientific revolution, Phys. Educ. 59 (2024) 055001. https://doi.org/10.1088/1361-6552/ad589c.
\bibitem{frolov}  V.P. Frolov, A. Zelnikov, Introduction to Black Hole Physics, Oxford University Press, Oxford, 2011.
\bibitem{zurek} W.H. Zurek, Entropy Evaporated by a Black Hole, Phys. Rev. Lett. 49 (1982) 1683–1686. https://doi.org/10.1103/PhysRevLett.49.1683.
\bibitem{prigogine} I. Prigogine, Introduction to Thermodynamics of Irreversible Processes, Wiley (1968).
\bibitem{stahler} S. W. Stahler and  Francesco Palla, The Formation of Stars, WILEY‐VCH Verlag GmbH \& Co. KGaA (2004).
DOI:10.1002/9783527618675
\bibitem{kippenhahn} R. Kippenhahn, A. Weigert, and A. Weiss, Stellar Structure and Evolution, Springer Heidelberg New York Dordrecht London, 2$^{\text{nd}}$ edition (2012).
DOI 10.1007/978-3-642-30304-3
\bibitem{suzuki} H. Suzuki, Neutrinos from Core-Collapse Supernova Explosions, Progress of Theoretical and Experimental Physics (PTEP), Vol. 2024, Issue 5, 05B101 (2024).
https://doi.org/10.1093/ptep/ptae056
\bibitem{reed} B. Reed and C.J. Horowitz, Total energy in supernova neutrinos and the tidal deformability and binding energy of neutron stars, Phys. Rev. D 102, 103011 (2020).
DOI: https://doi.org/10.1103/PhysRevD.102.103011

\end{thebibliography}
\end{document}